\documentclass[iop]{emulateapj}
\pdfoutput=1

\usepackage{natbib}
\usepackage{threeparttable}
\usepackage{amsmath}
\usepackage{amssymb}
\usepackage{graphicx}
\usepackage{subfigure}
\usepackage{threeparttable}

\newcommand{\HI}{\ensuremath{\mbox{\rm \ion{H}{1}}}}

\newcommand{\htwo}{\ensuremath{\mbox{H$_2$}}}
\newcommand{\msun}{\ensuremath{M_\odot}}

\newcommand{\pc}{\ensuremath{\mbox{pc}}}
\newcommand{\kms}{\mbox{km~s$^{-1}$}}

\newcommand{\cm}{\mbox{cm$^{-2}$}}

\newcommand{\co}[1]{\mbox{$^{#1}$CO}}

\newcommand{\av}{\ensuremath{\mbox{$A_{\rm V}$}}}

\newcommand{\counits}{\mbox{K km s$^{-1}$}}

\newcommand{\tmb}{\ensuremath{\mbox{$T_{\rm mb}$}}}
\newcommand{\tex}{\ensuremath{\mbox{$T_{\rm ex}$}}}
\newcommand{\nhtwo}{\ensuremath{\mbox{$N(\rm H_2)$}}}
\newcommand{\nh}{\ensuremath{\mbox{$N_{\rm H}$}}}

\shortauthors{Imara et al.}

\begin{document}

\title{X Marks the Spot: Nexus of Filaments, Cores, and Outflows in a Young Star-Forming Region}

\author{Nia Imara\altaffilmark{1}, Charles Lada\altaffilmark{1}, John Lewis\altaffilmark{1}, 
John H. Bieging\altaffilmark{2}, Shuo Kong\altaffilmark{3}, Marco Lombardi\altaffilmark{4}, 
Joao Alves\altaffilmark{5}}
\altaffiltext{1}{Harvard-Smithsonian Center for Astrophysics, 60 Garden Street, Cambridge, MA 02138}
\email{nimara@cfa.harvard.edu}
\altaffiltext{2}{Steward Observatory, The University of Arizona, Tucson, AZ 85719}
\altaffiltext{3}{Department of Astronomy, Yale University, 52 Hillhouse Avenue, New Haven, CT 06511}
\altaffiltext{4}{University of Milan, Department of Physics, via Celoria 16, I-20133 Italy}
\altaffiltext{5}{University of Vienna, T\"urkenschanzstrasse 17, 1880 Vienna, Austria}

\begin{abstract}
We present a multiwavelength investigation of a region of a nearby giant molecular cloud that is distinguished by a minimal level of star formation activity.  With our new \co{12}($J=2-1$) and \co{13}($J=2-1$) observations of a remote region within the middle of the California molecular cloud, we aim to investigate the relationship between filaments, cores, and a molecular outflow in a relatively pristine environment.  An extinction map of the region from \emph{Herschel Space Observatory} observations reveals the presence of two 2-pc-long filaments radiating from a high-extinction clump.  Using the \co{13} observations, we show that the filaments have coherent velocity gradients and that their mass-per-unit-lengths may exceed the critical value above which filaments are gravitationally unstable.  The region exhibits structure with eight cores, at least one of which is a starless, prestellar core.  We identify a low-velocity, low-mass molecular outflow that may be driven by a flat spectrum protostar.  The outflow does not appear to be responsible for driving the turbulence in the core with which it is associated, nor does it provide significant support against gravitational collapse.
\end{abstract}



\section{Introduction}
Observations show that the internal structures of giant molecular clouds (GMCs) are permeated with complex substructure, including networks of filaments on a wide range of spatial scales , molecular outflows, and newly forming stars.  In optical, submillimeter, and far-infrared images, filaments are frequently seen radiating from compact groups of young stellar objects (YSOs) and prestellar cores (e.g., Myers 2009; Andr\'e et al. 2010; Ward-Thompson et al. 2010; Andr\'e et al. 2014).  Millimeter observations have revealed that YSOs, in turn, are frequently associated with molecular outflows, which may drive a significant portion of turbulence in GMCs (e.g.,  Bally et al. 1996; Reipurth \& Bally 2001; Maury et al. 2009).   The ubiquity of such features suggests that outflows and filaments play critical roles in star formation (e.g., Margulis et al. 1988; Maury et al. 2009; Arzoumanian 2011; Hacar et al. 2016).  In particular, molecular line studies of nearby filamentary structure demonstrate that prestellar core formation depends on cloud kinematics and chemistry (e.g., Duarte-Cabral et al. 2010; Hacar \& Tafalla 2011; Arzoumanian et al. 2013; Kirk et al. 2013).

In this paper, we present results on the California molecular cloud (CMC), an excellent laboratory for studying the influence of environment, structure, and cloud evolution on the initial conditions of star formation.  At a distance of $450\pm 23$ pc (Lada et al. 2009, Lombardi et al. 2010), the CMC has a similar mass, size, and morphology to the Orion A molecular cloud (OMC).  However, the CMC has a much lower level of star formation activity than does the OMC, with roughly an order of magnitude fewer YSOs (Lada et al. 2009; Lada et al. 2010).  The most obvious area of star formation within the CMC is the young embedded cluster associated with the reflection nebula NGC 1579 (Andrews \& Wolk 2008), located at the southeastern edge of the cloud.  The cluster contains the B star Lk H$\alpha$ 101, the most massive star known in the CMC.  In dust extinction maps, one finds that the majority of the YSOs in the cloud are associated with regions of highest extinction (Lada et al. 2009; Harvey et al. 2013).  Most observed YSOs, in fact, are projected along a slender filamentary ridge in the south of the cloud (Lada et al. 2009; Harvey et al. 2013).  As one proceeds westward in the CMC, one encounters regions of fewer YSOs, sometimes only single objects, and regions with no apparent star formation activity at all.  Thus, not only is the CMC notable for its low level of star formation activity compared to more active GMCs in the Solar vicinity, but within the cloud itself there appears to be a gradient in the star formation rate (SFR) that is worthy of examination.  

The focus of this study is on a small region in the middle of the CMC, near $(l,b)=(162\fdg4, -8\fdg8)$, for which we present new \co{12}($J=2-1$) and \co{13}($J=2-1$) observations taken at the Arizona Radio Observatory.  These observations were motivated in part by a noteworthy feature previously observed in an extinction map: two filaments radiating from a high-density structure with two embedded massive cores.  As we will show, the high-density structure from which the filaments radiate also contains an outflow of molecular gas that is most evident in \co{12} emission.  The filaments themselves are most prominent in dust extinction.  Altogether, the structure as observed in extinction resembles an ``X,'' and we will henceforth refer to it as California-X, or Cal-X.  Near the outflow, infrared observations reveal the presence of two bright sources, one of which is a YSO candidate that could be driving the outflow.  While a number of studies have considered filaments feeding into clusters (e.g., Myers 2009; Schneider, N. et al. 2012; Andr\'e et al. 2016), there has been less of a focus on the role of filamentary structure in the formation of individual stars, as we observe in this interesting structure within the CMC.  All in all, Cal-X provides an ideal opportunity for investigating incipient star formation in a relatively pristine environment.  In particular, our observations enable us to examine the role of filamentary structure and molecular outflows in an early stage of stellar evolution.

The main goals of this paper are to characterize the physical properties of the various components of Cal-X---including the filaments, cores, and outflow---and to determine the kinematic relationship of these components with each other and with the YSO.  In Section \ref{sec:observations}, we describe the \co{12}(2-1),  \co{13}(2-1), and far-infrared \emph{Herschel Space Observatory} observations used for this study.  We also describe the properties of the YSO observed in this region.  Our methods and results on the physical properties of Cal-X are presented in Section \ref{sec:results}.  We discuss the implications of our results for the evolution of star formation in the CMC in Section \ref{sec:discussion}, and we summarize our conclusions in Section \ref{sec:conclusions}.

\section{Observations}\label{sec:observations}
\subsection{CO data}
Observations of the CMC were made with the Heinrich Hertz Submillimeter Telescope (HHT), a 10 meter submillimeter facility located at an elevation of 3200 m on Mount Graham, Arizona. The $J = 2-1$ lines of $^{12}{\rm C}^{16}{\rm O}$ (230.538 GHz; hereafter \co{12}) and $^{13}{\rm C}^{16}{\rm O}$ (220.399 GHz; hereafter \co{13}) were mapped over a $1\fdg 28 \times 0\fdg 48$ field centered at ({\it l,b}) = (162$\degr$, -8.85$\degr$) using a prototype ALMA Band 6 sideband separating receiver with dual orthogonal linear polarizations. \co{12} and \co{13} were simultaneously observed in the upper and lower sidebands, respectively. We use a 128 channel filterbank spectrometer with 0.25 MHz (0.34 km s$^{-1}$)  wide channels to collect data over  a range in V$_{\rm LSR}$ of $-20$ to 16 \kms~for both lines. The typical single sideband system temperature for these observations is about 200 K.  

Observations were conducted in November and December in 2012 as part of a larger HHT CO mapping survey of the CMC (e.g, Kong et al. 2015). The field was broken into 23 $10\arcmin \times 10\arcmin$ tiles on a $3\times8$ grid. Each tile is generated ``on-the-fly'' by raster scanning each tile field at a rate of $10\arcsec/{\rm sec}$ along the galactic longitude with 0.1 second sampling (smoothed with a 0.4 second window in post-processing). Each row in the raster was offset by $10\arcsec$ in galactic latitude. This sampling gives a native beam size of $34\arcsec$--$36\arcsec$, depending on the frequency. The data were calibrated using a standard chopper-wheel technique (Kutner \& Ulich 1981) to establish a temperature scale in antenna temperature, $T_A^*$, which was then corrected  to a main-beam temperature scale, $T_{\rm mb}$, using observations of the standard source W3(OH) (for details, see Bieging \& Peters 2011). The individual calibrated polarization maps are combined using the inverse-variance-weighted average and are stitched into a final map with overlapping regions weighted by their variances. The final \co{12} and \co{13} maps were convolved to a common beam FWHM size of $38\arcsec$. The final maps have a per pixel rms noise of 0.17~K/channel. 

\subsection{Far-infrared data}
Far-infrared observations of the CMC were taken with the \emph{Herschel Space Observatory} in the Auriga-California program (Harvey et al. 2013).  The observations were taken in the parallel mode of the PACS and SPIRE instruments (Griffin et al. 2010).  Additional details about the observing strategy may be found in Harvey et al. (2013) and Andr\'e et al. (2010).  After the methods of Lombardi et al. (2014), the \emph{Herschel} data products were pre-processed with the most recent version of the calibration files, using the \emph{Herschel Interactive Processing Environment} (HIPE; Ott 2010). A dust extinction map of the CMC map was created according the methods described in Lombardi et al. (2014).  To summarize, the dust emission, which is optically thin at the observing frequencies of \emph{Herschel} ($\lambda=160$--500 \micron), was modeled as a modified blackbody.  All the \emph{Herschel} data were first convolved to the 36\arcsec~beam size of the SPIRE 500 \micron~map---comparable to the beam size of the CO data---and then the optical depth was determined by fitting the observed spectral energy distribution according to the modified blackbody model.  

\begin{figure*}
\centering
\includegraphics[width=5.5in]{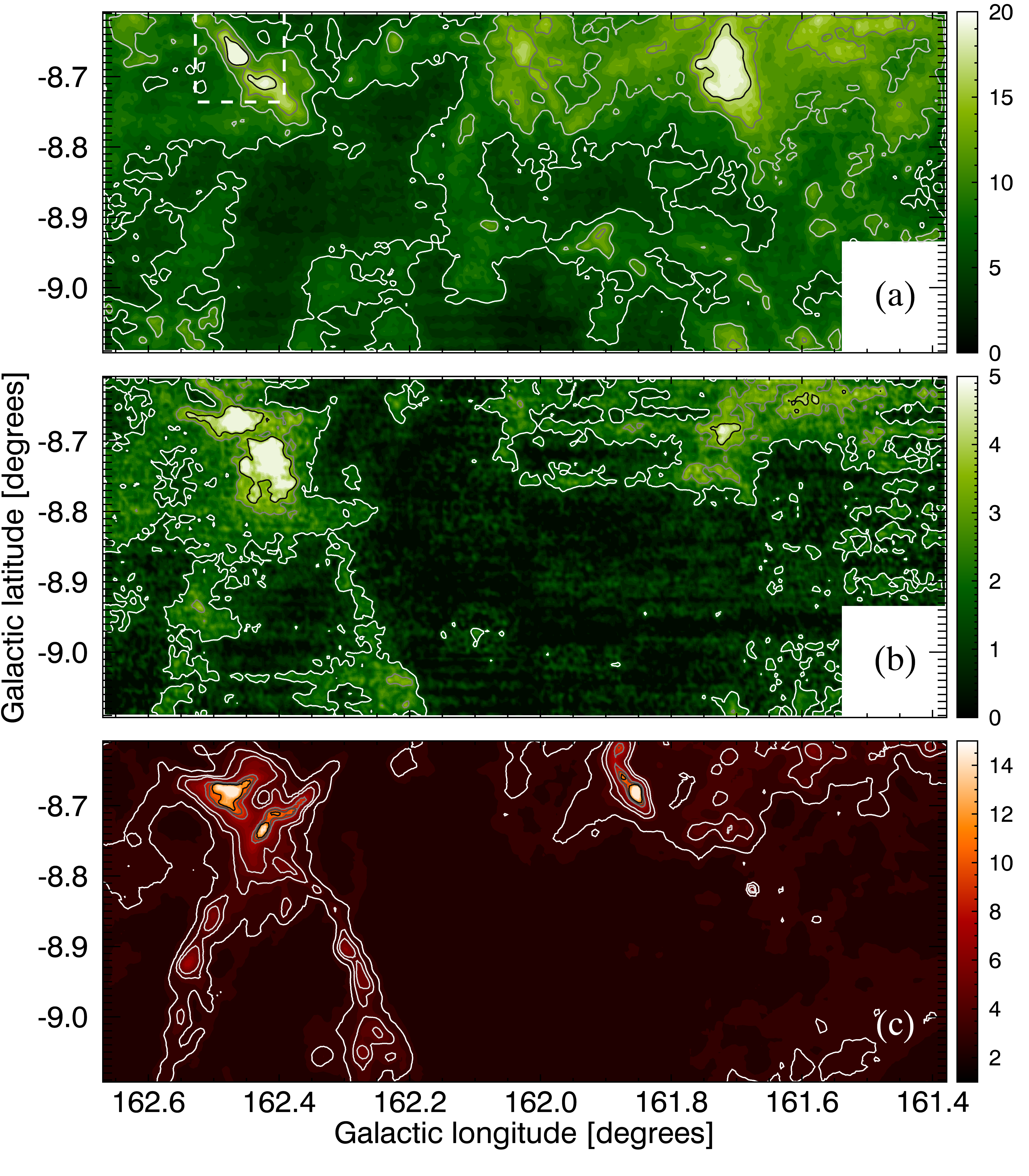} 
\caption{Integrated intensity maps of (a) \co{12} and (b) \co{13} with corresponding contours overlaid for clarity.  Both maps are integrated over the range $-8$ to 5 \counits, and the color bars are in units of \counits.  The grayscale \co{12} contour levels are 6, 10, 14, and 18 \counits.  The grayscale \co{13} contour levels are 1.5, 3, and 4 \counits.  (c) Dust extinction map derived from far-infrared \emph{Herschel} observations, with the color bar in units of visual magnitudes, \av.  The grayscale contour levels are 4, 6, 8, 16, 20, and 24 mag. \label{fig:maps}}
\end{figure*}

\section{Methods and Results}\label{sec:results}
\subsection{Dust Extinction and CO Intensity Maps}\label{sec:maps}
Figure \ref{fig:maps} displays the \co{12} and \co{13} intensity maps of the sub-region of the CMC, both integrated over the velocity range $-8$ to 5 \kms.  In the bottom panel of the figure is the dust extinction map of the region in units of visual extinction, \av, derived from \emph{Herschel} data, which calls attention to the presence of two prominent filamentary structures extending below approximately $-8\fdg 8$.  One can see that traces of the filamentary structures are apparent in the \co{13} map but are scarcely discernible in the \co{12} map.  However, in both the \co{12} and \co{13} channel maps, discussed below, the filamentary structures are more pronounced.  Nevertheless, dust is the more reliable tracer overall, since \co{12} tends to be optically thick at high column densities, while locally, \co{13} can also be optically thick or suffer from fractionation (e.g., Lada et al. 1994). A molecular tracer of higher densities, such as C$^{18}$O, is likely to recover more of the filamentary structure than either \co{12} or \co{13} (e.g., Figure 5 of Alves et al. 1999).  Finally, the dashed box in Figure \ref{fig:maps} draws attention to a notable feature in the north, centered at $(l,b)\approx (162\fdg 45, -8\fdg 7)$.  This feature is prominent in each map and particularly so in the \co{12} map.  In Section \ref{sec:outflow}, we show that this feature is an outflow.

\begin{figure*}
\centering
\includegraphics[width=5in]{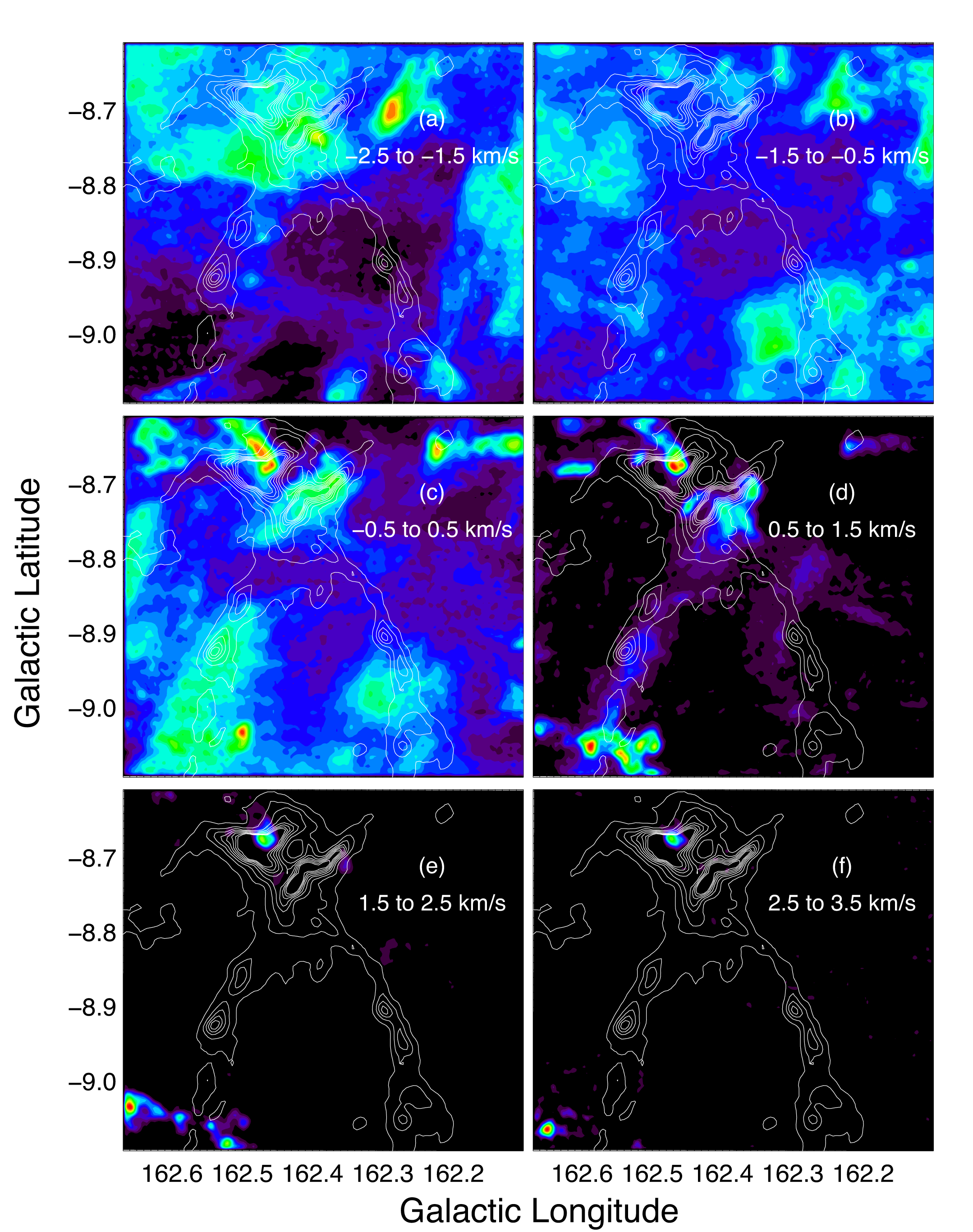} 
\caption{Channel maps of \co{12} emission, shown in color, integrated over channel widths of 1 \kms. The velocity range is shown on the right in each map.  Dust extinction contours, from $\av=2.2$ to 20 mag, are overlaid.  The color scale in each map goes from 0 \counits~to a maximum intensity of (a) 5.7, (b) 6.9, (c) 6.4, (d) 5.5, (e) 5.0 and (f) 2.9 \counits.   \label{fig:channel12}}
\end{figure*}

\begin{figure*}
\centering
\includegraphics[width=5in]{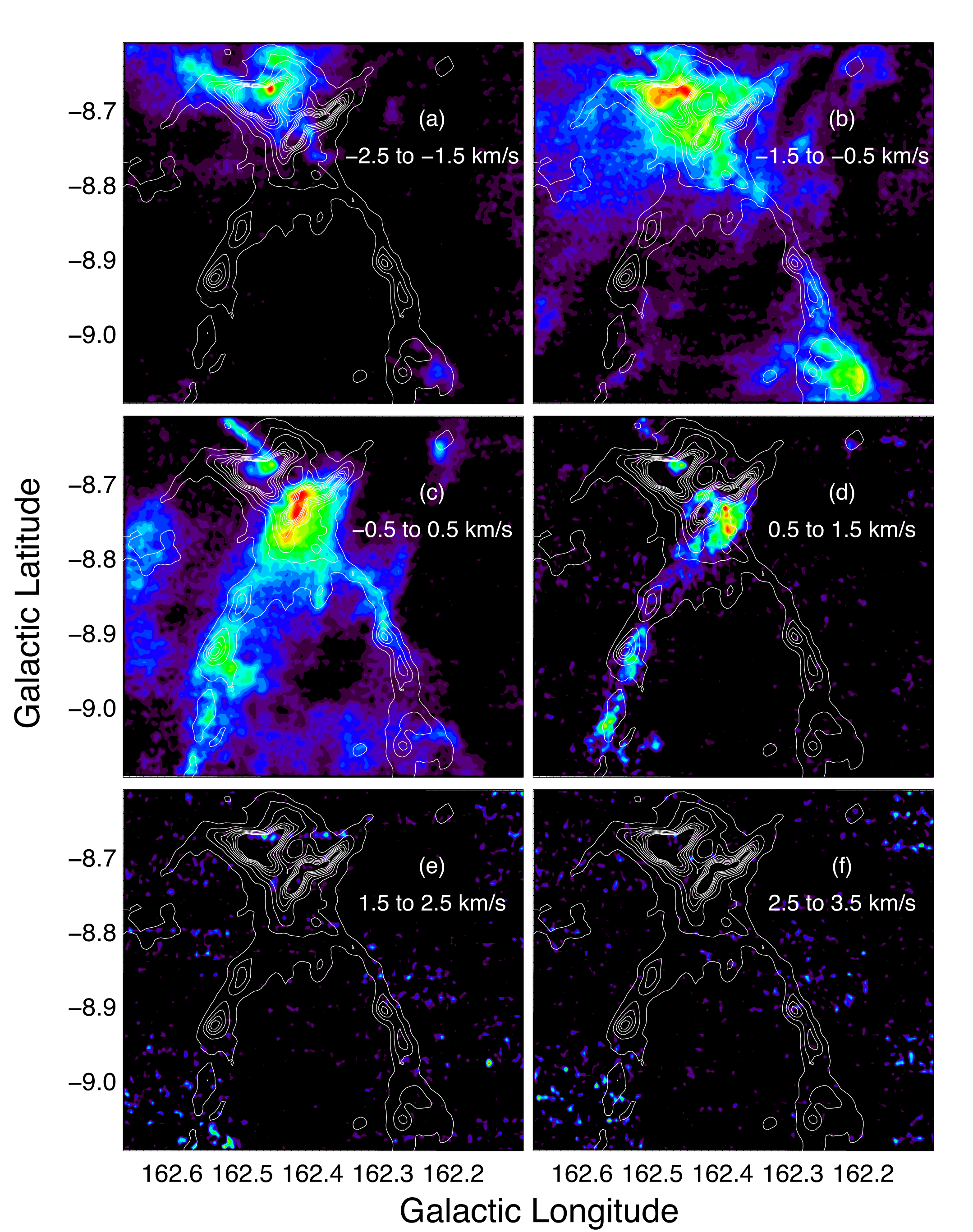} 
\caption{Channel maps of \co{13} emission.  Same as Figure \ref{fig:channel12}, except the color scale in each map goes from 0 \counits~to a maximum intensity of (a) 3.2, (b) 3.0, (c) 3.2, (d) 1.1, (e) 0.6 and (f) 0.5 \counits.  \label{fig:channel13}}
\end{figure*}

In Figures \ref{fig:channel12} and \ref{fig:channel13}, we show channel maps of the \co{12} and \co{13} emission, from $-2.5$ to 3.5 \kms, in bins of 1 \kms.  In these maps and the following, we now restrict our attention to the X-shaped feature, Cal-X, located at $l>162\fdg 1$, containing two filament-like structures in the south and the outflow in the north.  In panels (c) and (d) of the \co{12} channel maps, some of the filamentary structure becomes apparent in the velocity range $-0.5$ to 1.5 \kms.  The outflow becomes distinct from the rest of the emission at velocities greater than $\sim -0.5\kms$.  The \co{13} channel maps indicate that the two filamentary structures located at $b<-8\fdg 8$ have most of their emission at distinct velocities.  The emission from the west filament, located between $\sim 162\fdg 20 < l <162\fdg 33$, is concentrated within the velocity range of $-1.5$ to $+0.5$~\kms.  For the east filament, located between $\sim 162\fdg 48 < l <162\fdg 65$, most of the emitting material is moving at velocities $-0.5$ to $+1.5$ \kms.  To obtain a more detailed understanding of the filament kinematics, we analyze their spectra in \S\ref{sec:spectra} below. 

\begin{figure*}
\centering
\includegraphics[width=4.6in]{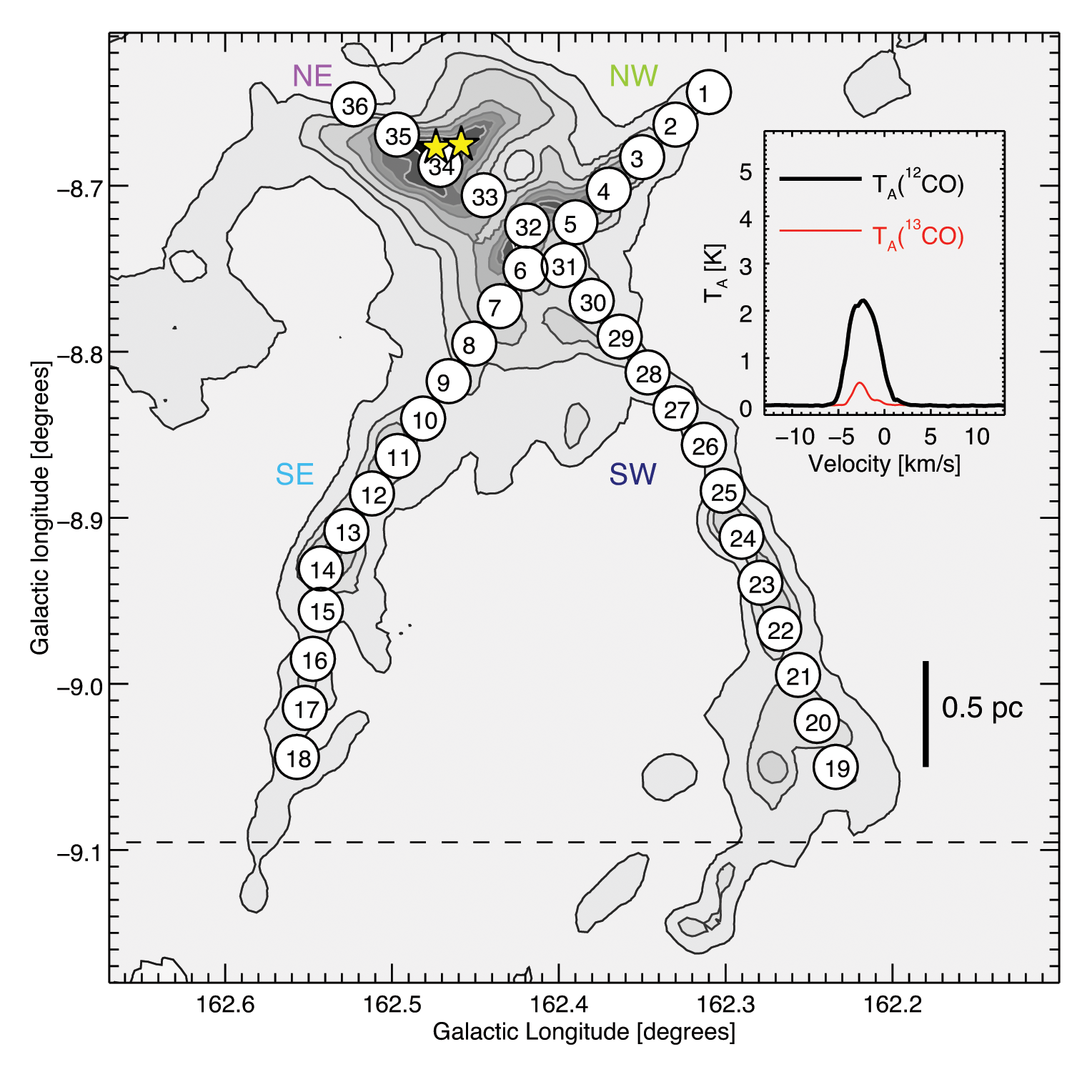} 
\caption{Key --- locations for spectra in the next figures.  The dust extinction map of Cal-X is shown in grayscale, with contour levels at 4, 6, 8, 12, 16, 20, 24, and 32 mag.  The colored labels identify the four axes defined in \S \ref{sec:maps}: northwest (NW), northeast (NE), southeast (SE), and southwest (SW).  The two stars mark the locations of two infrared bright sources, discussed in \S \ref{sec:ysos}.  The inset shows the average \co{12} and \co{13} spectra. \label{fig:multi_key}}
\end{figure*}

In the rest of this paper, our aim is to characterize the physical properties of the Cal-X substructures and determine how they might be related.  For convenience, we divide the region in the extinction map into four axes, as shown in Figure \ref{fig:multi_key}:  northwest (NW; positions 1--5), northeast (NE; positions 32--26), southeast (SE-fil; positions 6--18), and southwest (SW; positions 19--31).

\subsection{CO Spectra}\label{sec:spectra}
\begin{figure*}
\centering
\includegraphics[width=6.5in]{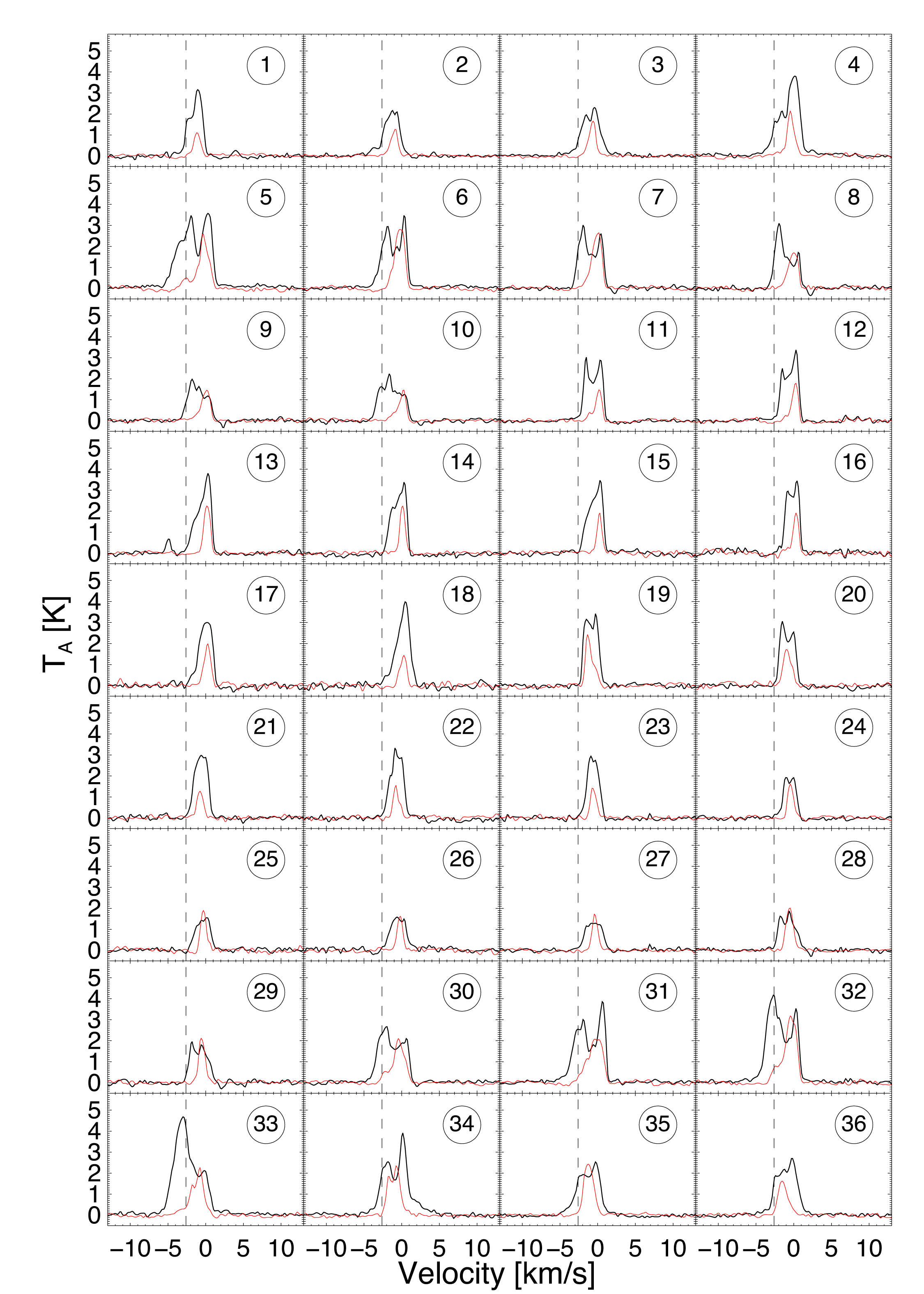} 
\caption{Spectra of the \co{12}~and \co{13}~lines (black and red lines, respectively) corresponding to the positions labeled in Figure \ref{fig:multi_key}.  Each spectrum was a extracted from a region having $5\times 5$ pixels, or $\sim 50\arcsec\times 50\arcsec$.  The dashed vertical line in each panel marks the systemic \co{13} central velocity of the entire Cal-X region, $-2.6$ \kms.   \label{fig:multispec}}
\end{figure*}

We characterized the kinematic properties of the substructures by analyzing the \co{12} and \co{13} spectra at each location labeled in Figure \ref{fig:multi_key}.  The average \co{12} and \co{13} spectra toward the entire region are displayed in the top right corner of Figure \ref{fig:multi_key}, and the \co{12} and \co{13} spectra corresponding to each location along the filaments are displayed in Figure \ref{fig:multispec}.  Although the \co{12} spectra appear structured with typically at least two peaks, the \co{13} spectra exhibit single peaks and are clearly more narrow than the \co{12} spectra.   We fitted a Gaussian to each individual \co{13} spectrum in Figure \ref{fig:multispec} using the MPFITFUN least-squares fitting procedure from the Markwardt IDL Library (Markwardt 2009).  The three parameters determined from the Gaussian fit at each location---the peak main-beam brightness temperature, \tmb, central velocity, $v_0$, and velocity dispersion, $\sigma$---are summarized in Table \ref{table1}.  The systematic \co{13} and \co{12} velocities of the entire Cal-X region as derived from the average  spectra toward the region are $-2.6$ and $-2.3$ \kms, respectively.  The average \co{13} and \co{12} linewidths are 1.9 and 3.5 \kms, respectively.  We note that the Gaussian fitting of the \co{13} spectra result in typical uncertainties on the velocity measurements, $\pm 0.01~\kms$, that are less than the channel width of the data (0.34 \kms).

\begin{figure*}
\centering
\includegraphics[width=4.6in]{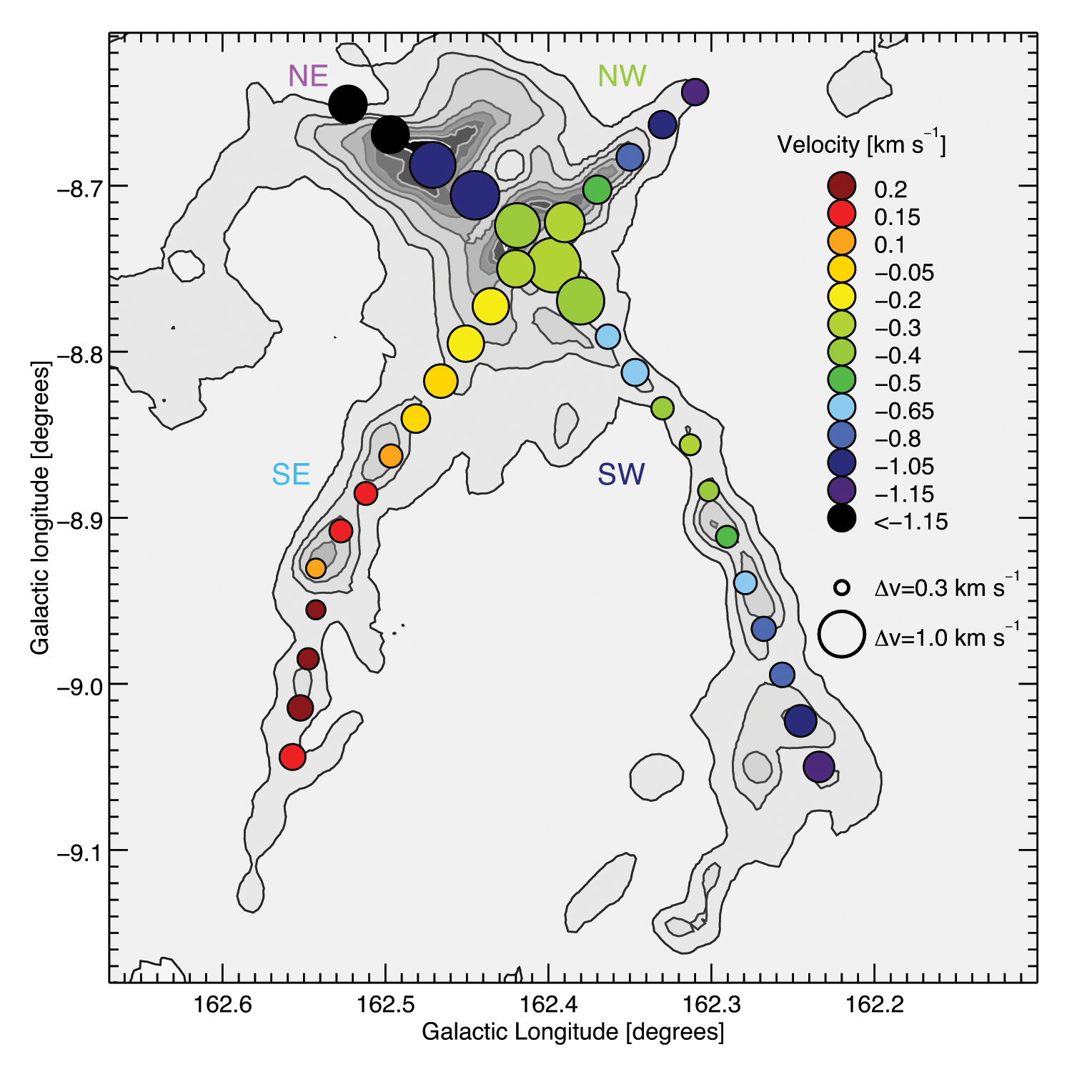} 
\caption{Results of the \co{13}~spectra fitting (circles) overlaid on the \emph{Herschel} extinction map.  The color and size of the circles represent the gas velocity and the gas velocity dispersion, respectively. \label{fig:multimap}}
\end{figure*}

The results of the Gaussian fitting are presented in Figure \ref{fig:multimap}.   The colors and sizes of the circles represent the line-of-sight velocity and linewidth, respectively.  Dispersions are fairly narrow and uniform along the filaments but broaden at their apex, near position 31.  The axes of Cal-X also converge dynamically near position 31 to a velocity of approximately $-0.35~\kms$, and each axis displays the presence of a velocity gradient.  We note, in particular, the velocity gradients along the lengths of the filamentary structures, which we denote as SE-fil and SW-fil.  These gradients could plausibly be due to accelerating gas flows along the filaments, in particular, either an inflow or an outflow of gas, depending on the relative orientation of the filaments along our line of sight (see \S\ref{sec:discussion1}).  The NE axis associated with the outflow is slightly blueshifted with respect to the other three substructures.  As Figure \ref{fig:multispec} shows, all four of the axes of Cal-X are redshifted with respect to the \co{13} systematic velocity of $-2.6~\kms$ of the entire region, indicated with a vertical dashed line, which is dominated by relatively low-density material.  Assuming a gas-to-dust ratio of $\nh/\av=1.87\times 10^{21}\cm~{\rm mag}^{-1}$ (Bohlin et al. 1978), the total mass of material in the region having $\av\ge 1$ mag is 2220 \msun, while the total mass of material with $\av\ge 4$ mag (the threshold extinction we use to define the extents of the filaments; see Section \ref{sec:filaments} below) is 950 \msun.

The \co{12} line profiles shown in Figure \ref{fig:multispec} are complex, suggesting multiple substructures that are moving at different velocities along the line-of-sight at many of the positions. The profiles of the rarer \co{13} isotope are generally simple, single-peak spectra that are well-characterized by single Gaussians.  Exceptions include positions 30--34, which indicate the presence of a second component and the possibility that \co{13} may be optically thick here.   A number of the \co{12} profiles have dips at velocities where there is a peak in \co{13} emission---for instance, positions 4, 8, 17, 20, 24, 32, and 34---signaling that the \co{12} emission is optically thick and self-reversed.

\begin{figure}
\epsscale{1.}
\plotone{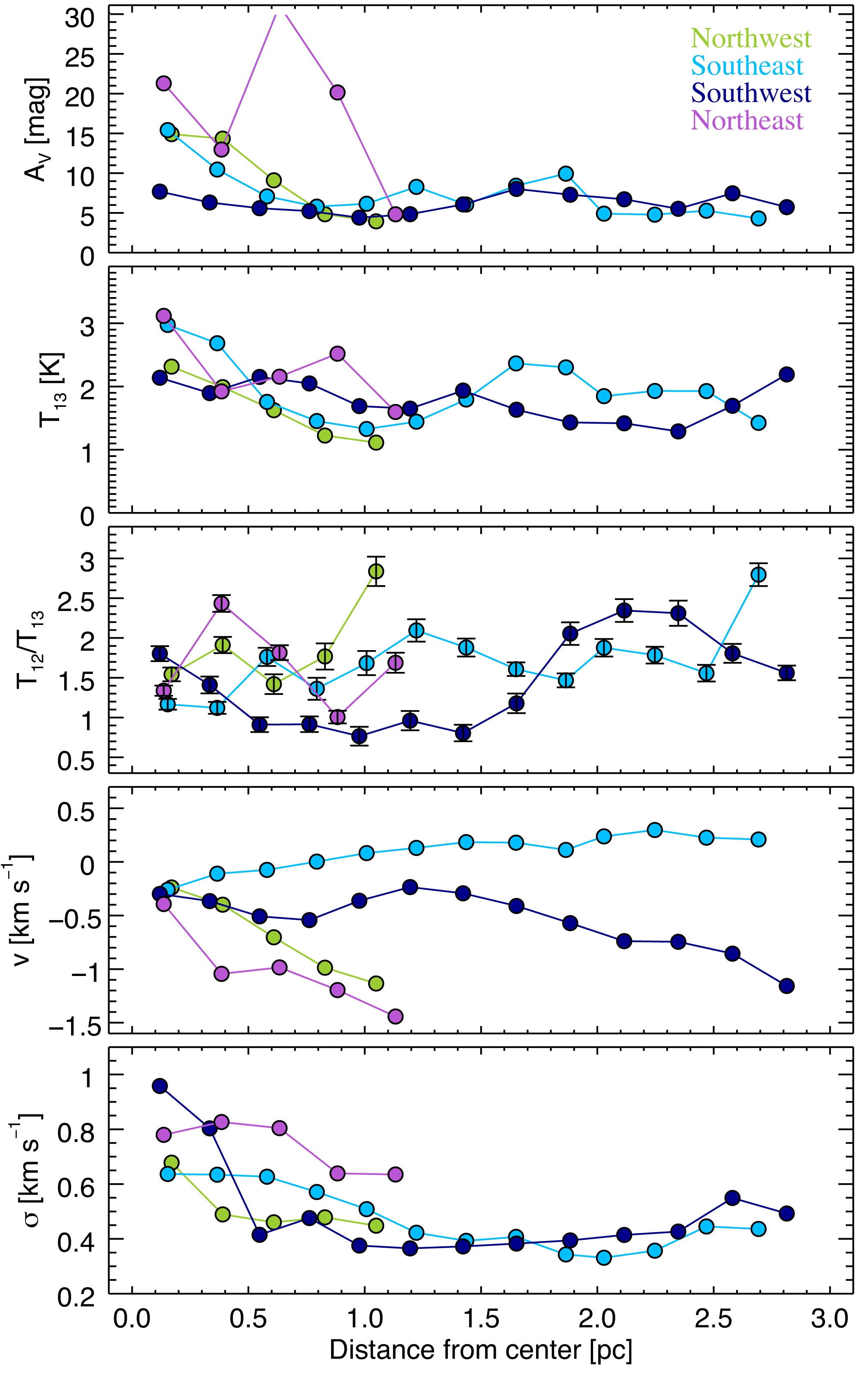} 
\caption{Correlations in extinction \av, \co{13} peak brightness temperature $T_{13}$, ratio of \co{12} peak brightness temperature to $T_{13}$, velocity $v$, and velocity dispersion $\sigma$ with distance from center of the filaments.  Only the error bars for $T_{12}/T_{13}$ are shown.    
The typical errors in \av~($\sim 0.02$ mag), $T_{13}$ ($\sim 0.01$ K), and $v$ and $\sigma$ ($\sim 0.005$ \kms) are smaller than the plot symbols.   \label{fig:cor1}}
\end{figure}

In Figure \ref{fig:cor1}, we present the parameters derived from the Gaussian fits to the \co{13} spectra as a function of distance along each axis of Cal-X.  For convenience, we calculate the distance from the nexus of the four axes, which we define to be the center of positions 5, 6, 31, and 32, located at $(l,b)=(162\fdg 406,-8\fdg736)$.  We note that, in the Dobashi et al. (2005) dark cloud catalog, this is near the location of cloud 1096 (see Table 7 in Dobashi et al. 2005).  In the first panel of Figure \ref{fig:cor1}, we also show \av~as a function of distance.  The \co{13} brightness temperature, \tmb, varies between 1--3 K and 1--2.5 K for the SE and SW filaments, respectively.  The ratio of the peak brightness temperatures, $T_{12}/T_{13}$, varies between $\sim 0.5$ and 3 K, substantially smaller than the fractional abundance on Earth of $\sim 89$ (Wilson \& Matteucci 1992), which demonstrates that the \co{12}(2-1) is optically thick in the line center.  The fourth panel in Figure \ref{fig:cor1} confirms the presence of a velocity gradient along each of the four axes that we see in Figure \ref{fig:multimap}.  In particular, the magnitudes of the velocity gradients along the filaments are roughly 0.1 $\kms~{\rm pc}^{-1}$ (SE-fil) and 0.2 $\kms~{\rm pc}^{-1}$ (SW-fil).  The gradients are roughly linear, but contain some departures from linearity that could be due to random motions or, as is the case with the NE axis, the presence of an outflow near the position where the \co{13} spectrum was taken.   The last panel in Figure \ref{fig:cor1} shows that the velocity dispersions along the four axes are highest close to the nexus point, near the outflow, and then decrease with distance from the nexus.

\begin{table}\centering
\caption{Kinematic properties of Cal-X.}\label{table1}
\begin{threeparttable}
\renewcommand{\arraystretch}{1.1} 
\begin{tabular}{lccccc}
\tableline\tableline
Position   &  $l$      & $b$       & Velocity  &  FWHM   &  \tmb   \\ 
           &(degrees)  &(degrees)  &(\kms)     & (\kms)  &  (K)    \\
\tableline
1           & 162.310   & $-8.644$   & $-1.12$    & 1.08   & 1.04   \\
2           & 162.330   & $-8.663$   & $-0.98$    & 1.13   & 1.21   \\
3           & 162.350   & $-8.683$   & $-0.71$    & 1.11   & 1.61   \\
4           & 162.370   & $-8.702$   & $-0.41$    & 1.16   & 1.98   \\
5           & 162.390   & $-8.722$   & $-0.24$    & 1.63   & 2.29   \\
6           & 162.420   & $-8.750$   & $-0.27$    & 1.53   & 2.92   \\
7           & 162.435   & $-8.773$   & $-0.13$    & 1.49   & 2.64   \\
8           & 162.451   & $-8.795$   & $-0.08$    & 1.49   & 1.74   \\
9           & 162.466   & $-8.818$   & $0.00 $    & 1.38   & 1.44   \\
10          & 162.481   & $-8.840$   & $0.09 $    & 1.17   & 1.37   \\
11          & 162.497   & $-8.863$   & $0.15 $    & 0.95   & 1.50   \\
12          & 162.512   & $-8.885$   & $0.17 $    & 0.94   & 1.78   \\
13          & 162.527   & $-8.908$   & $0.17 $    & 0.94   & 2.32   \\
14          & 162.543   & $-8.930$   & $0.13 $    & 0.81   & 2.25   \\
15          & 162.543   & $-8.955$   & $0.22 $    & 0.78   & 1.86   \\
16          & 162.547   & $-8.985$   & $0.27 $    & 0.87   & 1.83   \\
17          & 162.552   & $-9.014$   & $0.22 $    & 1.04   & 1.80   \\
18          & 162.557   & $-9.044$   & $0.20 $    & 1.06   & 1.33   \\
19          & 162.234   & $-9.050$   & $-1.12$    & 1.25   & 2.04   \\
20          & 162.245   & $-9.022$   & $-0.86$    & 1.30   & 1.64   \\
21          & 162.257   & $-8.995$   & $-0.75$    & 1.01   & 1.30   \\
22          & 162.268   & $-8.967$   & $-0.73$    & 0.99   & 1.36   \\
23          & 162.279   & $-8.939$   & $-0.58$    & 0.94   & 1.42   \\
24          & 162.290   & $-8.911$   & $-0.42$    & 0.90   & 1.61   \\
25          & 162.302   & $-8.884$   & $-0.31$    & 0.86   & 1.91   \\
26          & 162.313   & $-8.856$   & $-0.25$    & 0.86   & 1.66   \\
27          & 162.330   & $-8.834$   & $-0.36$    & 0.92   & 1.65   \\
28          & 162.347   & $-8.812$   & $-0.54$    & 1.11   & 2.02   \\
29          & 162.363   & $-8.791$   & $-0.51$    & 1.00   & 2.12   \\
30          & 162.380   & $-8.769$   & $-0.37$    & 1.94   & 1.86   \\
31          & 162.397   & $-8.748$   & $-0.30$    & 2.24   & 2.17   \\
32          & 162.419   & $-8.724$   & $-0.38$    & 1.84   & 3.07   \\
33          & 162.445   & $-8.706$   & $-1.02$    & 1.99   & 1.91   \\
34          & 162.471   & $-8.687$   & $-0.99$    & 1.88   & 2.18   \\
35          & 162.497   & $-8.669$   & $-1.19$    & 1.52   & 2.44   \\
36          & 162.523   & $-8.651$   & $-1.43$    & 1.54   & 1.60   \\
\tableline
\end{tabular}
\textbf{Note.} Properties derived from the fits to the \co{13} Gaussians in Figure \ref{fig:multispec}.  Column 1: position number corresponding to the key in Figure \ref{fig:multispec}.  Columns 2 and 3: position coordinates.  Columns 4--6: spectrum best-fit velocity, best-fit linewidth, and best-fit brightness temperature, respectively.   The mean uncertainty on all velocity measurements is $\pm 0.01$ \kms.  The mean uncertainty on \tmb~is 0.02 K.
\end{threeparttable}
\end{table}

\subsection{Physical Properties of Filaments and Cores}\label{sec:filaments}
Dust is the preferred tracer of total gas column density (e.g., Bohlin et al. 1978; Goodman et al. 2009), so to estimate the two filament masses and other related properties, we use the \emph{Herschel} extinction map.  We begin by assuming that the filaments include material having extinction of $\av\ge 4$ mag.  We found that setting the extinction boundary at $\av\ge 3.5$ mag includes too much extraneous material beyond the filaments.  On the other hand, in setting the boundary at $\av\ge 4.5$ mag, the continuity of the filamentary structure is lost.   The extinction map covers much more area than does the \co{13} map we use to measure the kinematic properties of the filaments.  Thus, we define the southernmost tip of SE-fil at $(l,b)=(162\fdg58,-9\fdg14)$, below position 18; the southernmost tip of SW-fil is located at $(l,b)=(162\fdg32,-9\fdg16)$. We define the northernmost parts of SE-fil and SW-fil at positions 9 and 28, respectively.  It is likely that the filaments extend even further northward---i.e., into the high-column-density region containing the NE and NW axes---but beyond this,  we have no way of separating them out from the high-column-density material into which they blend.

In Table \ref{table2} we list the sizes, masses, and number densities of the filaments.  Since the filaments may extend further north than we defined, our estimated lengths are possibly lower limits. We approximate the widths by taking the average of a number of measurements along the spines of the filaments.  Column densities are calculated assuming the standard Galactic dust-to-gas ratio
\begin{equation}\label{eq:gdr}
\frac{\nh}{\av} = 1.87\times 10^{21}~\cm~{\rm mag}^{-1},
\end{equation}
of Bohlin et al. (1978), where $\nh= N(\HI)+2\nhtwo$ is the total hydrogen column density. We then calculate the mass using
\begin{equation}\label{eq:mass}
M=\mu m_{\rm H} \nh S,
\end{equation}
where $\mu=1.36$ is the helium correction factor, $m_{\rm H}$ is the hydrogen mass, and $S$ is the surface area occupied by material having $\av\ge 4$ mag.  The resulting masses of SE-fil and SW-fil are roughly 126 and 154 \msun, respectively.  Table \ref{table2} also lists the estimated masses per unit length, which we discuss in \S\ref{sec:discussion}.  Under the hypothesis that the filaments are funneling gas to the high-density region in the north, we also estimate flow velocities and mass flow rates, $\dot{M}$, also to be discussed in \S\ref{sec:discussion1}.  All errors in Table \ref{table2} stem from the propagation of uncertainties in the \emph{Herschel} map and from the uncertainty in the distance to the California GMC, $d=450\pm 23$ pc.  We also include an uncertainty in the gas-to-dust ratio of 10\%.

The dust map in Figure \ref{fig:maps} reveals the presence of several compact, high-extinction subregions embedded throughout Cal-X.  We aim to identify cores, keeping two objectives in mind.  First, we want to select peaks in the extinction map that correspond to potentially star-forming material.  Here we draw on the argument that cloud material above the extinction of $\av\approx 7.3$ mag is most directly related to star formation activity (Lada et al. 2010).  This is similar to the threshold for core formation discussed by Andr\'e et al. (2010).  Second, we want to differentiate neighboring peaks from each other and from the background material.  With these criteria,  we identify six cores, defined as material enclosed within bounded contours of $\av\ge 7$ mag, embedded in SE-fil and SW-fil.  
Each of these cores contains a local maximum, i.e., an isolated peak in emission at a pixel in the map that is not connected to any other pixels at higher contour levels.

Since the cores in the northern region near the outflow are embedded in gas that has a much higher average extinction than the filaments, our criteria for defining their boundaries are somewhat stricter.  In this case, we identify two cores, delineating their boundaries at $\av=12$ mag.  At lower extinctions, the two cores start to blend together, but at 12 mag, the cores appear as distinct local peaks in the extinction map.  Contour maps of the total of eight cores we define in this way are shown in Figure \ref{fig:clump_key}.  The selection criteria are somewhat subjective, but they successfully isolate peaks in the extinction map and distinguish neighboring cores from one another.  Cores E and F in SW-fil are an exception, in that they appear blended together into one structure at 7 mag.  However, due to the prominent ``saddle'' in between these two local peaks in extinction, we decide to treat them as distinct objects and measure their properties separately. 

We compared our results with those of the core extraction program CLUMPFIND (Williams et al. 1994), an automated routine for analyzing hierarchical structure in observations of molecular clouds.  We used the version of the program for two-dimensional data sets.  The algorithm searches for peaks in emission located within a user-defined lowest level contour, and then follows the peaks until all the emission above the defined threshold can be assigned to specific clumps (Williams et al. 1994).  We applied CLUMPFIND to the \emph{Herschel} dust map with a range of lowest level contours as inputs, and found that at 7 and 12 mag levels, the algorithm extracts the same eight cores that we identify above. 

We estimate the effective radius of each core, $R_{\rm eff}=\sqrt{A_c/\pi}$, from its projected area, $A_c$, corresponding to continuous pixels having extinctions in excess of 7 or 12 mag. In calculating the core masses, to correct for the fact that they are embedded within regions of high extinction, we perform a background subtraction as follows.  For a core having a projected area, $A_c$, we calculate the mass within identical background areas having an extinction of 4 mag, $M_{\rm back}$.  For each core, we select at least two regions in the dust map that are near the core without having any overlapping emission with it or any other cores.  We define the average mass of these background areas as $M_{\rm back}$.
Next, we subtract $M_{\rm back}$ from the uncorrected mass contained within the core before subtraction, $M_{\rm uncor}$.  That is, we estimate the core mass as $M = M_{\rm uncor} - M_{\rm back}$. Similarly, in estimating the number density of a core, we subtract a contribution from its mean column density equivalent to $\av = 4$ mag.

In Table \ref{table3} we list the derived core sizes (ranging from 0.11 to 0.28 pc), the corrected masses (from $\sim 4$ to 88 \msun), the uncorrected masses (listed in parentheses), and number densities (from $\sim 1.2\times 10^4$ to $4.4\times 10^4$ cm$^{-3}$).

We note that the choice of contouring level effects the estimated masses, sizes, and other properties of the cores, such as the virial parameter, which we examine in \S\ref{sec:discussion}.  We performed additional calculations of the core masses and sizes at $\av=7\pm 1$ mag and $12\pm 1$ mag to get a sense of how much the estimates differ.  The average core mass at 6 mag is $\sim 1.3$ times greater than the estimated average at 7 mag, which in turn is $\sim 1.4$ times the average at 8 mag.  The average core size at 6 mag is 1.6 times larger than the average size at 7 mag, which is 1.3 times the average size at 8 mag.  In \S\ref{sec:discussion} we will discuss how the choice in contour level affects estimates of the virial parameter.

\subsection{Infrared source properties}\label{sec:ysos}
The two sources associated with core A, (A1 and A2 in Figure \ref{fig:clump_key}), were previously identified as YSOs by Harvey et al. (2013) and Broekhoven-Fiene et al. (2014).   Broekhoven-Fiene et al. (2014) performed SED modeling on \emph{Spitzer} observations of the CMC to determine the YSO classes and physical properties, which are summarized in Harvey et al. (2013).   Sources A1 and A2, corresponding to source 13 and 12 in Harvey et al. (2013),  are located at $(l,b)=(162\fdg 47, -8\fdg 68)$ and $(162\fdg 46, -8\fdg 68)$, respectively.  Broekhoven-Fiene et al. (2014) classified source A1 as a class I YSO and determined its bolometric luminosity to be 0.56 $L_\odot$.  The authors classified A2 as a flat spectrum source with a luminosity of 3.40 $L_\odot$.  Source A2 appears prominently in the near-infrared image in Figure \ref{fig:outflow}; in the plane of the sky, it is seen to be projected close to the peak of the blueshifted lobe of the outflow.  Source A1, too faint to be visible in the image, is projected close to the peak of the redshifted lobe of the outflow. 
 
\subsection{Outflow}\label{sec:outflow}
Inspection of the \co{12} spectra in the northwest region of Cal-X reveals an excess of emission at high velocities compared to other regions.  In particular, the \co{12} spectra in Figure \ref{fig:multispec} corresponding to positions 32--36 have surpluses of blue- and/or redshifted wing emission that suggest a molecular outflow.  However, there are other possible causes for this excess emission that must be considered before conclusively identifying the source as an outflow.  For instance, the emission could be due to small background or foreground clouds along the line of sight, or it may arise from gas clumps within the CMC that are moving at speeds markedly different from the average cloud velocity.  Because the presence of excess wing emission is not by itself sufficient to identify an outflow, we follow Margulis et al. (1988) in adopting more selective criteria, which are aided by the contour map and spectrum displayed in Figure \ref{fig:outflow}.

\begin{figure*}
\centering
\includegraphics[width=4.5in]{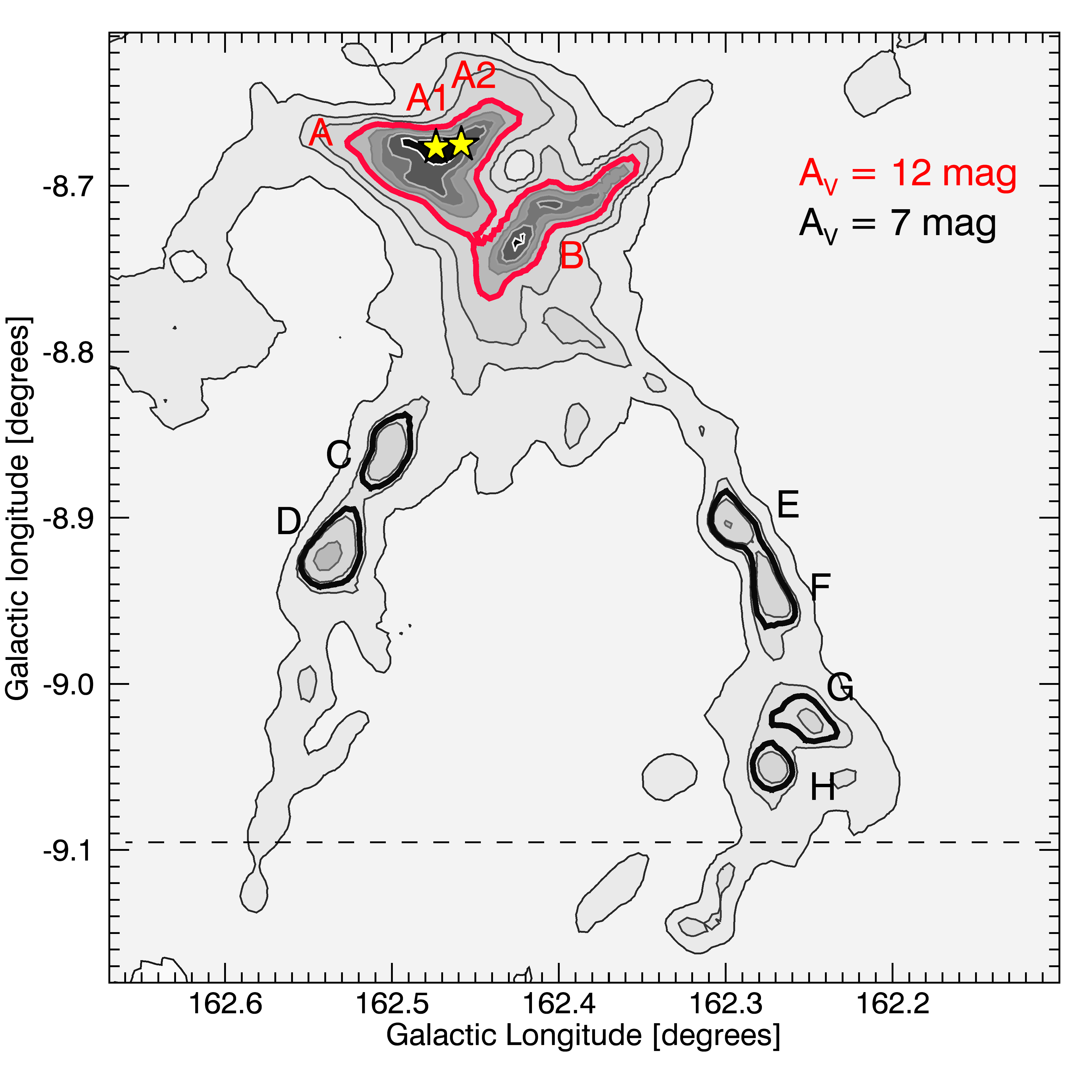} 
\caption{Key --- locations of cores whose properties are summarized in Table \ref{table3}. \label{fig:clump_key}}
\end{figure*}

First, our candidate outflow exhibits bipolarity.  Second, the shape and profile of its average \co{12} spectrum departs significantly from the average spectrum toward Cal-X as a whole, in which most subregions do not display excess emission.  Quantitatively, the average \co{12} linewidth of Cal-X is 3.5 \kms, whereas the average spectrum in the vicinity of the outflow exhibits velocities much greater than $(3.5~\kms)/2=1.75$ \kms~from the line center.  Third, two YSOs are close to the center of the bipolar outflow candidate and may be its driving source.  These three criteria are summarized in Figure \ref{fig:outflow}, which shows a near-infrared image of the field containing the sources, created  from \emph{JHK} observations taken at the Calar Alto Observatory in Spain.  The image, kindly provided by Carlos  Rom\'an-Z\'u\~niga, is overlaid with a contour map of the blue- and redshifted lobes of the outflow, as well as the average \co{12} spectrum toward the region in the inset.  Only the brighter of the two YSOs (A2 in Figure \ref{fig:clump_key}) is apparent in the near-infrared image.

To derive the masses of the outflow lobes using the \co{12} emission, we first assess the \co{12} optical depth. Analysis of the line ratios in \S\ref{sec:spectra} indicates that the \co{12} is likely to be optically thick toward the line center.  We estimate the average ratio of line intensities, $T_{12}/T_{13}$, toward the outflows as a function of velocity.  In the wings, where \co{13} emission becomes undetectable, we calculate upper limits to the ratio, by setting the minimum values of $T_{13}$ to the 1-$\sigma$ rms noise level, $0.1$ K.  We find that the maximum value of $T_{12}/T_{13}$ toward the outflow is $\sim 7$, indicating a moderate opacity for the \co{12} line.  Nevertheless, we estimate the masses of the outflow lobes according to the following method, under the assumption that the \co{12} line is optically thin.  Thus, we are likely to underestimate the optical depth, and therefore the mass.

For each velocity channel in the blue- or redshifted lobe, we calculate the \co{12} optical depth, $\tau_{12}$, using 
\begin{equation}\label{eq:tau}
\tau_{12}(v) = -\ln\left[ 1 - \frac{\tmb(\co{12})}{J(\tex) - J(T_{\rm bg})}   \right],
\end{equation}  
where $J(T) = \frac{h\nu/k}{\exp(h\nu/kT)-1}$ is the radiation temperature (Rohlfs \& Wilson 2004), and $\tmb(\co{12})$, the \co{12} main-beam temperature, depends on velocity.   Again, these calculations assume that the \co{12} is optically thin.

Next, we assume local thermodynamic equilibrium (LTE) conditions to estimate the \co{12} column density, $N_{12}$, in each channel using
\begin{equation}\label{eq:column}
\begin{split}
N_{12} &=2.39 \times 10^{14} (\tex + 0.93) \\
      &\times \int \frac{\tau_{12}(v)}{1 - \exp(-11.06/\tex)}~dv~\cm,
\end{split}
\end{equation}
where $\tex$ is the excitation temperature, and velocities over which we integrate are in units of \kms~(e.g., Scoville et al. 1986).  For the blueshifted (redshifted) lobe, we integrate over the velocity range $-10$ to $-4$ \kms~(1 to 7 \kms).   Finally, we assume an abundance ratio of $[\co{12}]/[\htwo]=10^{-4}$ to calculate the total mass per channel,
\begin{equation}\label{eq:mass_flow}
M_{\rm flow} =\mu m_{\rm{H}_2}  N_{12} A_{\rm flow}\left(\frac{[\rm{CO}]}{[\htwo]}\right)^{-1}
\end{equation}
where $A_{\rm flow}$ is the area of the outflow, $m_{\rm{H}_2}$ is the mass of an \htwo~molecule, and $\mu=1.36$ is the helium correction factor.  The resulting masses, calculated for   three different values of the unknown excitation temperature ($\tex=10,20,30$ K), are listed in Table \ref{table4}.  For $\tex = 10$ K, the masses for the blue- and redshifted lobes of the outflow are $\sim 0.08$ and 0.07 \msun, respectively.


\begin{table*}\centering
\caption{Physical properties of filaments.}\label{table2}
\begin{threeparttable}
\setlength{\tabcolsep}{12pt}
\begin{tabular}{lcccccc}
\tableline\tableline
Region       &  Size      &  Mass   & $\langle \rm{Density}\rangle$ &   $\langle M/L \rangle$  &  Infall Velocity  &  $\dot{M}$   \\ 
             &(pc$\times$pc) &(\msun)  &($\times 10^4$ cm$^{-3}$)   & (\msun~pc$^{-1}$)  &  (\kms) &  (\msun~Myr$^{-1}$) \\
\tableline
Filament East   & $2.3\times 0.3$  & $127\pm 16$  &  $1.8\pm 0.2$   &  $54.6\pm 6.8$  & $0.20$   &  $10.7\pm 1.3$   \\
Filament West   & $2.4\times 0.4$  & $154\pm 20$  &  $1.6\pm 0.2$   &  $65.0\pm 8.4$  &  $0.59$  &  $38.1\pm 4.9$   \\
\tableline
\end{tabular}
\textbf{Note.} Properties calculated for $\av \ge 4$ mag.
\end{threeparttable}
\end{table*}

Because the \co{12} may not be optically thin, the outflow masses we determine above are likely to be lower limits. It would be preferable to use \co{13} to determine the masses, since the molecule is more likely to be optically thin.  However, while we did not detect \co{13} toward the outflow, we noted earlier that the maximum value of $T_{12}/T_{13}$ toward the outflow is $\sim 7$.  Thus, if we assume that $T_{13}=T_{12}/7$ everywhere in the outflow, we can use analogous forms of equations (\ref{eq:tau}) and (\ref{eq:column}) to calculate $\tau_{13}(v)$ and $N_{13}$ and determine upper limits to the masses.  Estimated in this way, replacing $N_{12}$ with $N_{13}$ in equation (\ref{eq:mass_flow}), and assuming an abundance ratio of $[\co{13}]/[\htwo]=2\times 10^{-6}$ (e.g., Dickman 1978; Frerking et al. 1982), the masses of the outflow lobes are up to $\sim 7$ times higher than the estimates based on the assumption of optically thin \co{12}.

\section{Discussion}\label{sec:discussion}
\subsection{Filaments and cores}\label{sec:discussion1}
The continuous velocity gradients observed along the two Cal-X filaments in Figure \ref{fig:multimap} might be explained by one of a number of different processes.  The main possibilities are gas inflow (collapse), outflow (expansion), or rotation.  Differential rotation seems unlikely, since it would rip apart each filament on the order of a crossing time.  Taking an average 3D velocity dispersion of a filament, $\sigma_{3\rm D}=\sqrt{3}\sigma \approx \sqrt{3}(0.49~\kms) = 0.84$ \kms, and a width of $\sim 0.3$ pc, we get a crossing time of $t_{\rm cross}\approx 0.36$ Myr.   If SE-fil and SW-fil are at least as old as the YSOs in the region, $\sim 1$--3 Myr, they are probably gravitationally bound or pressure confined. Since there is no expanding source, such as an HII region or molecular outflow, at the nexus of the two filaments, the outflow scenario also seems unlikely.  The molecular outflow associated with the YSOs is located northward of the filaments and does not appear to be spatially or dynamically related to them.  We therefore suggest that the smooth velocity structure is best explained by a global infall of gas toward the center of the system.  Considering that the dominant gravitational potential well in Cal-X is the high-extinction clump containing the massive cores A and B, gravity could readily explain the velocity gradients, if all the gas is falling toward that region.  In this scenario, the redshifted filament SE-fil is in the foreground, relative to the center of the system, while the blueshifted SW-fil is in the background.

To further support the claim of infall, we can use conservation of energy arguments to check whether the observed velocities along the filaments are consistent with gravitational infall toward the massive clump.  A test mass starting from rest at infinity and falling toward a clump of mass $M$ and radius $R$ will reach a speed of $v=(2GM/R)^{1/2}$ by the time it arrives at the surface of the clump.  In convenient units, $v= 0.93~\kms~(M/100~\msun)^{1/2}(R/\rm{pc})^{-1/2}$.  We calculate this expected $v$ for a range of total clump masses and radii above a certain \av~threshold.  For $4\le \av\le 13$ mag, $9342\gtrsim M \gtrsim 963~\msun$, $5\gtrsim R\gtrsim 1$ pc, and $4\gtrsim v\gtrsim 3$ \kms.  Thus, the hypothesis of gravitational infall is plausible since these velocities are much larger than the observed velocities along the filaments (see Figure \ref{fig:multimap}).  It is likely that a confluence of other physical effects---including external pressures on the filaments, turbulence, and the fact that the filaments are not point masses but extended structures---are working together to oppose gravitational attraction toward the clump, thus slowing down the infall speeds.

The estimates of the mass-per-unit-lengths of SE-fil and SW-fil are, respectively, $M/L=54.6\pm 6.8$ and $65.0\pm 8.4~\msun~\pc^{-1}$.  Andr\'e et al. (2010) suggested that cores can form from filaments that  have attained the thermal critical mass-per-unit length, $(M/L)_{\rm crit} = 2c_s^2/G$ (Ostriker 1964), beyond which filaments are gravitationally unstable.  For a $\sim 7$ K isothermal filament,\footnote{The typical observed \co{12} temperature along the two filaments is $T_{\rm obs}\approx 3$ K. The corrected temperature from the Rayleigh-Jeans approximation, $T_{\rm cor}\approx 7$ K, is derived by solving $T_{\rm obs}=[J_\nu(T_{\rm cor})-J_\nu(T_{\rm cmb})](1-{\rm e}^{-\tau_\nu})$ for $T_{\rm cor}$, assuming $\tau_\nu\gg 1$.  $J_\nu(T)=(h\nu/k)/({\rm e}^{h\nu/kT}-1 )$, and $T_{\rm cmb}=2.725$ K is the cosmic microwave background temperature.} 
the corresponding sound speed is $c_s = \sqrt{kT/\mu m_{\rm H_2}} = 0.15~\kms$, and $(M/L)_{\rm crit} = 10.4~\msun~\rm{pc}^{-1}$.  The observed values of SE-fil and SW-fil are higher by a factor of $\sim$ 5--6, suggesting that they are thermally supercritical.  However, magnetic fields may be a source of support against gravitational collapse (Fiege \& Pudritz 2000; Beuther et al. 2015; Kirk et al. 2015).  In addition, turbulent motions may help stabilize the filaments against gravity.  Following Peretto et al. (2014), we consider an effective critical mass-per-unit-length that depends on the average velocity dispersion along the filaments, $(M/L)_{\rm eff} = 2\bar{\sigma}^2/G$.  Taking $\bar{\sigma}= 0.4~\kms$ (see Table \ref{table1}), we calculate $(M/L)_{\rm eff}=74~\msun~\pc^{-1}$.  Since the ratio of the effective to the observed $M/L$ is $\sim 0.7$--0.9, this suggests that the filaments are subcritical or, at best, marginally critical.  This is equivalent to saying that the filaments are supported by a gas with an effective temperature defined by the velocity dispersion: $\mu m_{\rm H_2}\bar{\sigma}^2/k=53$ K.  As it turns out, nevertheless, the extinction map of Cal-X reveals the presence of a number of overdense structures.

In \S\ref{sec:filaments}, we identified eight cores embedded in Cal-X, six of which are embedded within the filaments SE-fil and SW-fil.  In theory, a core is considered to be prestellar if it is starless and self-gravitating (Andr\'e et al. 2000; Di Francesco et al. 2007; K\"onvyes et al. 2015).  Such prestellar cores are expected to develop into protostars.  With the exception of core A, the cores we identified are starless.  To assess whether they are also self-gravitating, we estimate the virial parameter, $\alpha$, for each core according to (Bertoldi \& McKee 1992)
\begin{equation}
\alpha_{\rm vir}\equiv \frac{M_{\rm vir}}{M}=\frac{5\sigma^2 R}{GM},
\end{equation}
where  $M_{\rm vir}$ and $M$ are the virial and observed masses of the core, respectively.  An object is self-gravitating if $\alpha_{\rm vir}\le 2$ (Bertoldi \& McKee 1992).  In this way, using the corrected masses of the cores, we identify core B, with $\alpha_{\rm vir}\approx 1.5$, as the only self-gravitating core.   The other cores have virial parameters in the range 3--7 and are therefore likely to be pressure confined.  We note that if $\alpha_{\rm vir}$ is calculated using the uncorrected masses (i.e., not background-subtracted), we find that cores A and D are also marginally self-gravitating, with $\alpha_{\rm vir}\approx 1.9$ for both.  The virial analysis may not be valid for core A, however, since it may be affected by the outflow. 

As discussed in \S \ref{sec:filaments}, the choice of contour level affects the estimated mass of cores.  However, changing the extinction threshold (of 7 and 12 mag) by $\pm 1$ mag results in only slight changes to the typical core mass and size, by factors less than 2.  We made additional estimates of the virial parameter using these alternate values of the core mass and size and found that our results are robust.

In their study of a filamentary system composed of three \emph{Spitzer} dark clouds, Peretto et al. (2014) similarly conclude that the velocity gradients along the filaments are due to infalling gas toward the center of the system where two massive cores are situated.  Peretto et al. (2014) decompose this system, called SDC13, into four filaments, with lengths ranging from 1.6 to 2.6 pc, and widths ranging from 0.2 to 0.3 pc, similar to the Cal-X filaments.  The SDC13 filaments, however, are more massive ($\sim 195-256~\msun$) than SE-fil and SW-fil, and their mass-per-unit-lengths are likely critical or supercritical.  Three of the SDC13 filaments are each embedded with four to fice cores each, whereas the Cal-X filaments have two and four cores.  Perhaps the greater number of cores in the SDC13 filaments are a consequence of them having values of mass-per-unit-length that are closer to supercritical.

\begin{table*}\centering
\caption{Physical properties of cores.}\label{table3}
\begin{threeparttable}
\setlength{\tabcolsep}{18pt}
\begin{tabular}{lccccc}
\tableline\tableline
Core    & $\ell$      &  \emph{b} & $R_{\rm eff}$  &  [Uncorrected] Mass   & $\langle \rm{Density}\rangle$   \\ 
        & (Degrees) & (Degrees) & (pc)            & (\msun) &  ($\times 10^4$ cm$^{-3}$)   \\ 
\tableline
A  & 162.47 & -8.69    & 0.29       & [113.4] $84.4\pm 10.9$    &  $2.7\pm 0.3$   \\
B  & 162.41 & -8.72    & 0.16       & [65.3]  $58.3\pm 6.9$     &  $4.0\pm 0.5$   \\
C  & 162.50 & -8.86    & 0.14       & [11.6]  $ 6.2\pm 1.4$     &  $1.5\pm 0.2$   \\
D  & 162.54 & -8.92    & 0.17       & [18.2]  $10.8\pm 2.3$     &  $1.6\pm 0.2$   \\
E  & 162.29 & -8.90    & 0.13       & [9.9]   $ 5.5\pm 1.2$     &  $1.7\pm 0.2$   \\
F  & 162.27 & -8.95    & 0.14       & [10.7]  $ 5.7\pm 1.3$     &  $1.5\pm 0.2$   \\
G  & 162.25 & -9.02    & 0.13       & [8.4]   $ 4.0\pm 1.0$     &  $1.2\pm 0.2$   \\
H  & 162.27 & -9.05    & 0.11       & [6.8]   $ 3.5\pm 0.8$     &  $1.7\pm 0.2$   \\
\tableline
\end{tabular}
 \textbf{Note.} For cores A and B, properties calculated for $\av\ge 12$ mag.  For cores C -- H, properties calculated for $\av\ge 7$ mag.  Column 5 lists the masses of cores after background subtraction, as well as the uncorrected masses, in square brackets, before background subtraction.  See \S \ref{sec:filaments}.
\end{threeparttable}
\end{table*}

\subsection{Outflow energetics}
Protostellar outflows, particularly in cluster-forming groups, are thought to play a critical role in driving molecular cloud turbulence (Bally et al. 1996; Reipurth \& Bally 2001; Knee \& Sandell 2000; Maury et al. 2009).  How significant is the impact on its surroundings of one outflow driven by a single protostar?  In order to investigate the outflow energetics and determine its possible relationship to turbulence in Cal-X, we estimate the outflow momentum flux and mechanical luminosity and compare them to the relevant quantities determined for core A, with which the outflow is spatially associated.   To begin, we use the \co{12} mass estimates of the outflow provided in Table \ref{table4} to calculate its momentum and kinetic energy,
\begin{equation}\label{eq:momentum}
\begin{split}
P_{\rm flow} &= M_{\rm flow} v_{\rm char}, \\
KE_{\rm flow} &= \frac{1}{2}M_{\rm flow} v_{\rm char}^2.
\end{split}
\end{equation}
We determine $v_{\rm char}$, the characteristic velocity of an outflow lobe, from the intensity-weighted mean absolute velocity summed over the appropriate velocity range,
\begin{equation}\label{eq:vchar}
v_{\rm char} = \frac{v_{\rm obs}}{\cos i}=\frac{1}{N_{\rm pixels} \cos i}\sum_{\rm pixels} \frac{\sum \tmb(v) |v-v_0|}{\sum \tmb(v)}
\end{equation}
where $v_{\rm obs}$ is the observed velocity and $v_0=-2.6~\kms$ is the \co{13} systemic velocity of the entire region.  In calculating equation (\ref{eq:vchar}), we assume $v_{\rm char} = v_{\rm obs}$.   But since we do not know the inclination angle $i$ between the line of sight and the flow axis, our estimates of $v_{\rm char}$ may be underestimated.  For instance, if $i\le 60^\circ$, $v_{\rm char}$ would be underestimated by as much as a factor of two.  In this way, we calculate the characteristic velocities of the blueshifted and redshifted outflows to be 3.4 and 5.8 \kms, respectively.  These values, as well as $P_{\rm flow}$ and $\rm{KE}_{\rm flow}$ are summarized in Table \ref{table5}.  The total radially directed momentum of the outflow is 0.67 \msun~\kms, implying that by the time the outflow mass has reached a velocity of 1 \kms~from the line center, it will have swept up 0.67 \msun~of material.  This is only about $\sim 0.8\%$ of the mass of core A, suggesting that roughly 130 similar generations of outflows would be necessary to support core A against gravitational collapse.  How likely is this scenario?

Core A appears to be marginally self-gravitating.  Based on its \co{13} velocity dispersion and mass, we estimate $\alpha_{\rm vir}\approx 2.4$ (or, using the uncorrected mass, $\alpha_{\rm vir}\approx 1.9$).  To determine whether the force exerted by the outflow on the core can hinder its collapse, we estimate the momentum flux (or ``force'') of the outflow and compare this to the force required to keep the core in hydrostatic equilibrium.  To determine the momentum flux, $F_{\rm mom}=P_{\rm flow}/t_{\rm dyn}$, we estimate the dynamic timescale of the two outflow lobes as $t_{\rm dyn}=l_{\rm flow}/v_{\rm char}$, where $l_{\rm flow}$ is the length of a given outflow lobe.  We do not consider the inclination $i$ of the outflow, but we note that $t_{\rm dyn}$ depends on $i$ via both $v_{\rm char}$ and $l_{\rm flow}=l/\sin i$, where $l$ is the observed projected length on the sky.  The projected lengths of the blue- and redshifted lobes are roughly 0.73 and 0.30 pc, respectively, corresponding to $t_{\rm dyn}\approx 0.2$ and 0.05 Myr. The resulting momentum flux of the blue- and redshifted lobes are 1.4 and 7.8 \msun~\kms~Myr$^{-1}$ (see Table \ref{table5}).  Based on the upper limit estimate of the outflow mass from \co{13} observations (see \S\ref{sec:outflow}), the total momentum flux of the outflow may be up to $\sim 7$ times higher.  Following Maury et al. (2009), the combined force necessary to balance gravity at radius $R$ in core A is
\begin{equation}
F_{\rm grav}=4\pi R^2 P_{\rm grav}= \frac{GM^2}{2R^2},
\end{equation}
where $P_{\rm grav}=GM^2/8\pi R^4$ is the hydrostatic pressure for a spherical shell of radius $R$ encompassing a mass $M=2\sigma^2 R/G$. Thus, $F_{\rm grav}\approx 187~\msun~\kms~{\rm Myr}^{-1}$.  The total momentum flux of the outflow is a  factor of $\sim 20$ smaller than $F_{\rm grav}$, suggesting that the outflow is too weak to provide significant support against collapse in core A.  Even if the outflow mass were at the estimated upper limit, the total momentum flux would still be less than  $F_{\rm grav}$ by a factor of $\sim 3$.

To assess the possible influence of the outflow on turbulence in core A, we approximate the mechanical luminosity of the outflow, that is, the amount of kinetic energy that it delivers to the surrounding ISM, using $L_{\rm mech} = \rm{KE}_{\rm flow}/t_{\rm dyn}$.  Not taking into account inclination, the total mechanical luminosity of the outflow lobes is $\sim 4\times 10^{-3}~L_\odot$.  The upper limit is $\sim 28\times 10^{-3}~L_\odot$.
Compared to the 68 molecular outflows compiled by Lada (1985), who found that $L_{\rm mech}$ scales with the luminosity of the driving sources, the outflow and YSOs in Cal-X have very low luminosities.
 We compare this value of $L_{\rm mech}$ to the rate of turbulent energy dissipation in core A.  Following Mac Low (1999) and Maury et al. (2009), we use the one-dimensional \co{13} velocity dispersion determined for core A, $\sigma\approx 1.8~\kms$, to estimate the turbulent energy dissipation rate,
\begin{equation}
L_{\rm turb} = \frac{\frac{1}{2} M\sigma^2}{R/\sigma},
\end{equation}
where $M=84~\msun$ and $R=0.29$ pc are the mass and size of the core.  We find that $L_{\rm turb}\approx 0.14~L_\odot$, about a factor of 5--35 times greater than the mechanical luminosity of the outflow.  These are lower limits, since much of the mechanical luminosity will be radiated away in shocks.  Thus, it appears that the outflow is not a dominant contributor to the observed turbulence in the region.

\begin{figure*}
\centering
\includegraphics[width=4.5in]{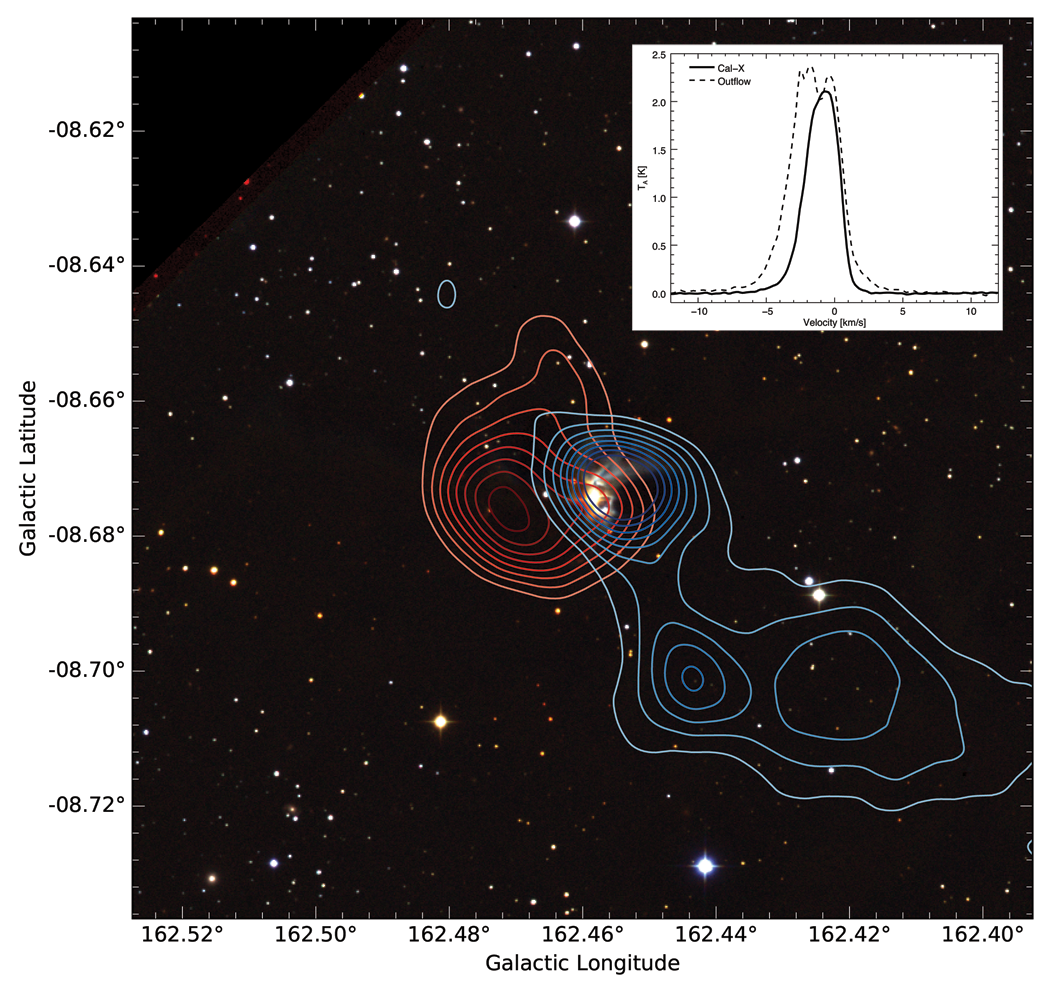} 
\caption{Near-infrared image of a region of the CMC, created from \emph{JHK} observations, overlayed with a contour plot of the outflow.  The blueshifted (redshifted) emission has been integrated over the velocity range $-10$ to $-4~\kms$ (1 to 7 \kms).  The contour levels for both blue- and redshifted emission go from 2 to 9 \counits, in steps of 1 \counits. The inset shows the average \co{12} spectra toward the entire Cal-X region (solid line) and in the vicinity of the outflow (dashed line).  \label{fig:outflow}}
\end{figure*}

\subsection{Origin of Cal-X}
The two most massive cores, A and B, located northward of the filaments, have a combined mass of $\sim 143~\msun$.  In \S\ref{sec:filaments}, we estimated the mass inflow rates of the SE-fil and SW-fil to be $\dot{M}\approx 10.7$ and 38.1 \msun~Myr$^{-1}$, respectively.  Supposing that the filaments have been funneling gas to the massive cores at a constant rate, and assuming they are the main suppliers of gas to the cores (as opposed to accretion, for example), it would have taken roughly 2.9 Myr for the cores to acquire their current total mass.  This is comparable to the median age of $2\pm 1$ Myr of YSOs observed in GMCs (e.g., Covey et al. 2010).  It is not obvious, however, whether the filaments formed before or after the sub-region embedded with the massive cores or if the regions are coeval.  If the filaments existed first, then perhaps the present mass of the cores is primarily due to the infall of gas from the filaments.  This scenario seems unlikely, though, given that all of the cores within SE-fil and SW-fill are starless and only one is self-gravitating, which suggests that the filaments could not be older than the region containing the protostars.  Thus, we suggest that the filaments were born together with \emph{or} after the massive region to the north, which may have attained the bulk of its present mass by some mechanism that predated the existence of the filaments.

In the filament, velocity gradients result from an overall inflow of mass, and if the cores are moving at the same infall velocity as the rest of the material in the filaments, then the filaments may be ``raining cores'' into the high-mass clump in the north.  In this scenario, based on their observed projected distances from the northernmost ends of the filaments, (defined in \S\ref{sec:filaments}), the cores in SE-fil and SW-fil may take $\sim 2$--5 Myr and $\sim 1$--3 Myr, respectively, to fall into the high-mass clump.  The free-fall times of these cores, $t_{ff}=\sqrt{3\pi/32 G\rho}$, are roughly 0.4--0.5 Myr.  The ``actual'' timescale for a core to traverse the length of a filament, $t_{\rm real}$, depends on the inclination of the filament, as $t_{\rm real} = t_{\rm obs} \cos i/\sin i$, where $t_{\rm obs} $ is the timescale estimated from observed filament properties.  In order to arrive at timescales as low as the core free-fall times of 0.4--0.5 Myr, the filaments would have to be inclined by at least $\sim 70$--$75^\circ$.  Thus, unless the filaments are highly inclined with respect to our line of sight, or unless the collapse of these cores is being slowed by some mechanism such as turbulent support, one might expect the cores to begin forming stars \emph{before} (and if) they ultimately ``rain'' into the high-mass clump.

\begin{table}\centering
\caption{Outflow velocity ranges and areas.}\label{table4}
\begin{threeparttable}
\setlength{\tabcolsep}{3pt}
\begin{tabular}{lccccc}
\tableline\tableline
Lobe     &  Velocity  &  Area      &  Mass            & Mass            &  Mass    \\
         &  Range     &            & [$\tex = 10$ K]  & [$\tex=20$ K]   & [$\tex=30$ K] \\
         &    (\kms)  & (pc$^2$)   & ($10^{-2}\msun$) &  ($10^{-2}\msun$) &  ($10^{-2}\msun$) \\
\tableline
Blue     &  -10 -- - 4   & 0.60      & 8.3   &  2.7  & 1.6 \\
Red      &   1 -- 7      & 0.65       & 6.9   &  2.2  & 1.3 \\
\tableline
\end{tabular}
\textbf{Note.} Outflow masses for three values of \tex, for the velocity ranges given in column 2.  The masses are determined from \co{12} observations, assuming the \co{12} is optically thin.
\end{threeparttable}
\end{table}

\begin{table*}\centering
\caption{Outflow energetics.}\label{table5}
\begin{threeparttable}
\setlength{\tabcolsep}{12pt}
\begin{tabular}{lccccc}
\tableline\tableline
Lobe     & $v_{\rm char}$  & Momentum     & Momentum Flux            & Kinetic Energy   & Mechanical Luminosity   \\ 
         &  (\kms)        & (\msun~\kms)  &  (\msun~\kms Myr$^{-1}$) & ($10^{42}$ erg)   & ($\times 10^{-3}~L_\odot$)    \\
\tableline
Blue     &   3.4      & 0.28   & 1.4  & 9.6  & 0.4 \\
Red      &   5.7      & 0.39   & 7.8  & 22   & 3.6  \\
\tableline
\end{tabular}
\textbf{Note.} Energetics of the blue- and redshifted lobes of the outflow.
\end{threeparttable}
\end{table*}

\section{Summary and Conclusions}\label{sec:conclusions}
In this paper, we have presented \co{12}($J=2-1$) and \co{13}($J=2-1$) observations of a region of the California GMC with a low level of star formation activity.  Mapped in extinction, this $\sim 0\fdg4\times 0\fdg6$ region ($\sim 3.1$ pc $\times~6.3$ pc at the distance to the CMC) exhibits two prominent filaments, each just over 2 pc in length, radiating from a high-column-density feature embedded with an outflow driven by one or two isolated YSOs.  Because of its resemblance to an ``X'' in the extinction map, we dub this region Cal-X.  Our main results are summarized as follows.

\begin{enumerate}
\item We examine the \co{13} spectra along the filaments and find that they possess velocity gradients along their lengths, with magnitudes of about 0.1 and 0.2 $\kms~{\rm pc}^{-1}$ for the southeast and southwest filaments, respectively.  The masses-per-unit-length of SE-fil and SW-fil are, respectively, $M/L=55$ and $65~\msun~\pc^{-1}$.  This exceeds the critical thermal mass-per-unit-length, $\sim 10~\msun~\pc^{-1}$ (assuming a sound speed of 0.095 \kms), above which filaments become gravitationally unstable (Ostriker 1964).  However, if we consider that non-thermal motions may support the filaments against gravity, the effective critical mass-per-unit-length defined by the velocity dispersion of the filaments yields $(M/L)_{\rm eff}=74~\msun~\pc^{-1}$, suggesting that the filaments are subcritical or marginally critical.

\item Based on the observed coherent velocity structure of the filaments and their spatial proximity to the large gravitational potential well to their north, we suggest that the velocity gradients are best explained by a northward infall of gas.  The combined mass inflow rate of the filaments is $\sim 49~\msun~\pc^{-1}$.

\item We identify eight cores embedded throughout Cal-X, six of which are associated with the filaments.  The two largest, most massive cores are located in the high-extinction feature north of the filaments; one is a starless, prestellar core; the other is associated with two bright infrared sources.  The background-subtracted core masses range from $\sim 3.5$--84 \msun, and 7 out of 8 of them have virial parameters $\alpha_{\rm vir}=3$--7, implying that they are confined by pressure, not gravity.

\item From the \co{12} data, we identify a low-velocity, low-mass outflow in the north of Cal-X.  The outflow is spatially coincident with the highest mass core (84 \msun), and it is associated with the two infrared-bright objects that may be its driving source. The blue- and redshifted lobes of the outflow have projected characteristic velocities of 3.4 and 5.7 \kms, momentum fluxes of 1.4 and 7.8 \msun~\kms~Myr$^{-1}$, and mechanical luminosities of $0.4\times 10^{-3}$ and $3.6\times 10^{-3}~L_\odot$.  The outflow appears to be too weak to contribute significantly to the turbulence in the core or to prevent the core from undergoing gravitational collapse.

\item Based on the available observations, we propose that the Cal-X filaments were born at the same time or after the massive region to their north, which may have attained the bulk of its present mass by some mechanism that predated the existence of the filaments. Therefore, given the presence of young YSOs in Cal-X, the filaments are likely to have an age of at least $2\pm 1$ Myr.

\end{enumerate}

\acknowledgements
We thank Carlos Rom\'an-Z\'u\~niga for providing us with the near-infrared image in Figure \ref{fig:outflow}.  We thank an anonymous referee and Phil Myers for helpful suggestions that improved earlier drafts of this paper.  
The Heinrich Hertz Submillimeter Telescope is operated by the Arizona Radio Observatory, which is part of Steward Observatory at the University of Arizona, and is supported in part by grants from the National Science Foundation.
N. Imara is supported by the Harvard-MIT FFL Postdoctoral Fellowship.

\end{document}